\documentclass{article}
\usepackage{ijcai24}

\usepackage{times}
\usepackage{soul}
\usepackage{bm}
\usepackage{url}
\usepackage[hidelinks]{hyperref}
\usepackage[utf8]{inputenc}
\usepackage[small]{caption}
\usepackage{graphicx}
\usepackage{amsmath}
\usepackage{amsthm}
\usepackage{booktabs}
\usepackage{algorithm}
\usepackage{algorithmic}
\usepackage[switch]{lineno}
\usepackage{enumitem}
\usepackage{subcaption}
\usepackage{multirow}
\usepackage{amssymb}

\title{Shadow-Free Membership Inference Attacks: \\Recommender Systems Are More Vulnerable Than You Thought
}

\author{
Xiaoxiao Chi$^1$
\and
Xuyun Zhang$^{1}$\footnote{Corresponding Author. Contact xuyun.zhang@mq.edu.au or Hongsheng.Hu@data61.csiro.au}\and
Yan Wang$^{1}$\and
Lianyong Qi$^2$\and
Amin Beheshti$^1$\and\\
Xiaolong Xu$^{3}$\and
Kim-Kwang Raymond Choo$^{4}$\and
Shuo Wang$^{5}$\and
Hongsheng Hu$^{6*}$\\
\affiliations
$^1$Macquarie University\\
$^2$China University of Petroleum (East China)\\
$^3$Nanjing University of Information Science and Technology\\
$^4$The University of Texas at San Antonio\\
$^5$Shanghai Jiao Tong University\\
$^6$CSIRO's Data61\\
}

\begin{document}

\maketitle

\begin{abstract}
Recommender systems have been successfully applied in many applications. Nonetheless, recent studies demonstrate that recommender systems are vulnerable to membership inference attacks (MIAs), leading to the leakage of users' membership privacy. However, existing MIAs relying on shadow training suffer a large performance drop when the attacker lacks knowledge of the training data distribution and the model architecture of the target recommender system. To better understand the privacy risks of recommender systems, we propose shadow-free MIAs that 
directly leverage a user's recommendations for membership inference. Without shadow training, the proposed attack can conduct MIAs efficiently and effectively under a practice scenario where the attacker is given only black-box access to the target recommender system. The proposed attack leverages an intuition that the recommender system personalizes a user's recommendations if his historical interactions are used by it. Thus, an attacker can infer membership privacy by determining whether the recommendations are more similar to the interactions or the general popular items. We conduct extensive experiments on benchmark datasets across various recommender systems. Remarkably, our attack achieves far better attack accuracy with low false positive rates than baselines while with a much lower computational cost.

\end{abstract}

\section{Introduction}
Recommender systems aim to accurately predict and suggest items or contents for users, which are widely applied in many real-world applications~\cite{zhang2019deep}, such as e-commerce sites~\cite{zhou2018micro}, healthcare domains~\cite{narducci2015recommender}, and social platforms~\cite{tang2016recommendations,wu2019neural,fan2019graph}. The success of recommender systems is largely attributed to the increasing availability of large-scale data generated by or associated with end users. The data often contain user profiles or behavioral information like age, gender, and shopping preference, thereby requiring strong protection on user privacy in terms of many recently issued laws and regulations like GDPR~\cite{rosen2011right} and CCPA~\cite{pardau2018california}. However, recent studies~\cite{zhang2021membership,wang2022debiasing,zhu2023membership} demonstrate that recommender systems are vulnerable to membership inference attacks (MIAs)~\cite{shokri2017membership}, where an attacker can infer the membership privacy of a user, i.e., distinguish member users whose data was used for training the recommender system from non-member users of the model. MIAs can directly reveal the privacy of a user in recommender systems. For example, if an attacker identifies that a user's data has been used for training a healthcare recommender system for treatment plans, the attacker can infer the user is a patient with a high chance. In addition, MIAs can be the foundations of other types of attacks, e.g., data extraction attacks~\cite{carlini2019secret,carlini2021extracting}.
Because of such abilities, MIAs have been widely used for measuring the privacy risks of machine learning models~\cite{carlini2022membership,hu2022membership}. 

To implement MIAs, a common approach is \textit{shadow training}, where the attacker trains a shadow model to mimic the behavior of the target model. Because the shadow model is trained by the attacker, the attacker can collect the features of members and non-members of the shadow model, which can be used for training a binary classifier as the attack model. Because the shadow model mimics the target model, the attack model trained on the shadow model will also work on the target model, which is often referred to as the attack transferability~\cite{salem2019ml}. Following the pipeline of shadow training, existing studies on MIAs targeting recommender systems~\cite{zhang2021membership,wang2022debiasing,zhu2023membership} are highly effective in inferring whether an individual's data was used to train a recommender system or not, e.g., the work in~\cite{zhang2021membership} shows that MIAs with the use of shadow training can achieve attack accuracy near $100\%$ against an item-based collaborative filtering recommender system trained on the Movielens-1M dataset. As a pre-requisite of shadow training, existing MIAs on recommender systems have a key assumption that the attacker owns the prior knowledge about the training data distribution and the model architecture of the target model to ensure the attack transferability.~\cite{zhang2021membership,wang2022debiasing,zhu2023membership}. With this assumption, the shadow model can be expected to behave similarly to the target model. However, it is often difficult for an attacker to obtain the prior knowledge in practice and the assumption fails to hold, since recommender systems are usually deployed under MLaaS (Machine Learning as a Service) environments~\cite{ribeiro2015mlaas} and only black-box access is available to the public. Besides, a shadow dataset from the same distribution of the training dataset is very difficult to satisfy completely. While existing works~\cite{zhang2021membership,wang2022debiasing,zhu2023membership} claim their abilities to generate a shadow dataset through querying the target model or using marginal distributions of training data, they often simply set a part of training dataset aside as the shadow dataset in empirical evaluations. In addition to this limitation, MIAs with shadow training are computationally expensive, requiring considerable computational resources for training both the shadow model and the attack model. Therefore, how to address these two drawbacks is still a challenge in existing MIAs with shadow training.

From a defence perspective, the above-mentioned impractical assumption may give a sense of security in recommender systems: They can stay safe from MIAs as long as the attacker does not have the assumed prior attack knowledge. Indeed, existing works~\cite{zhang2021membership,wang2022debiasing,zhu2023membership} have tried to train a shadow model without the prior attack knowledge. Instead, they make use of a different shadow model architecture and a shadow dataset from a different distribution, but the resultant attack model suffers from a large performance drop, sometimes even to a level of randomly guessing. This demonstrates failure to have the assumption hold will disable MIAs with shadow training. However, this sense of security is false as existing MIAs on recommender systems fail to take the special aspects of recommender systems into consideration. Recommender systems usually recommend personalized items to members, as their historical interactions were used for training the model. Such personalized items, through meticulously chosen by the recommender system, are similar to members' historical interactions. However, for non-members, recommender systems usually recommend generally popular items, because the model has not seen their historical interactions~\cite{sedhain2014social}.

In regard of this, in this paper we propose shadow-free MIAs without any process of shadow training with only black-box access to the model. The attack intuition is to examine whether a user's recommendations are more similar to his historical interactions or general popular items. If the recommendations are more similar to historical interactions, the attacker can infer the user as a member, and infer the user as a non-member otherwise. The challenge here is how to obtain general popular items of a recommender system, which serves as an important reference for the similarity comparison. To solve this challenge, we skillfully leverage the characteristics of recommender system scenarios. Specifically, an attacker can generate an empty user account without historical interactions with the target recommender system. Then, the attacker can collect the recommendations of the empty user and consider them as general popular items. This implementation is easy for the attacker to achieve in practice, e.g., creating a new account in Amazon. Being lightweight, the newly proposed attacks can efficiently and effectively conduct MIAs against recommender systems.

Our contribution is summarized as follows:
\begin{itemize}[leftmargin=*]

\item This paper is the \textit{first} to investigate shadow-free MIAs against recommender systems. The proposed new attack is lightweight and can effectively infer the membership privacy of a user with only black-box access to the recommender system.

\item Extensive experiments are conducted on three benchmark datasets across various recommender systems and compared with representative baseline attacks. Experimental results demonstrate that the newly proposed attacks achieve far better performance than baselines in terms of attack effectiveness, reliability, and attack efficiency.

\item The source code of the shadow-free MIAs is released at \url{https://github.com/XiaoxiaoChi-code/shadow-free-MIAs.git}, which creates a new tool for measuring the privacy vulnerability of recommender systems and sheds light on the design of future defense methods.

\end{itemize}

\section{Related Work}
\paragraph{Recommender Systems.} Recommender systems enrich user experiences by predicting and suggesting items within a vast array of content. Among various recommendation algorithms~\cite{burke2002hybrid,chen2017attentive}, traditional collaborative filtering recommendation algorithm~\cite{koren2009matrix} is the main stream, which aims to recommend items to users based on their preferences and behaviors of other users with similar tastes. In recent years, deep learning techniques have been widely applied to recommender systems. Advanced deep learning based recommender systems leverage neural networks to model complex patterns and representations of user-item interactions. Techniques such as autoencoders~\cite{sedhain2015autorec}, neural collaborative filtering~\cite{he2017neural}, recurrent neural networks~\cite{hidasi2015session}, and long short-term memory networks~\cite{liu2018stamp,zhou2019deep} have demonstrated significant success in improving recommendation accuracy and addressing challenges associated with sparse and high-dimensional data.

\paragraph{Membership Inference Attacks.} MIAs aim to infer whether a data sample was used to train a target model or not. The work~\cite{shokri2017membership} firstly investigates MIAs on classification models. Later works~\cite{hayes2017logan,song2019auditing,he2020segmentations,he2021node} further investigate the feasibility of MIAs on other types of models such as image generative and segmentation models. Given the simplicity of the definition, MIAs have been considered as a standard metric for measuring the privacy of machine learning models~\cite{carlini2022membership,ye2022enhanced,song2021systematic}. A few works~\cite{zhang2021membership,zhu2023membership,wang2022debiasing} have investigated the membership privacy risks on recommender systems and demonstrated that MIAs are less effective if an attacker does not know the prior knowledge of the training data distribution and the model architecture of target recommender system. This phenomenon is rational because existing MIAs rely on shadow training. Once the shadow recommender system is not similar enough to the target one, the attack model built on the shadow recommender system cannot transfer well to the target one. In addition, shadow training is usually associated with high computational costs for training the shadow and attack models. In this paper, we fulfill this research gap by proposing shadow-free MIAs, which can efficiently and effectively conduct the MIAs without shadow training.

\section{Methodology}
In this section, we first introduce the threat model of MIAs in recommender systems. Then, we introduce shadow-based MIAs and analyze their limitations, which serve as a motivation for proposing more powerful and practical MIAs. Last, we introduce our proposed method for shadow-free MIAs.
\subsection{Threat Model}
In this paper, we study MIAs under the black-box settings, i.e., we assume an attacker can only query the target recommender system and obtain the recommended items. Following previous works~\cite{zhang2021membership,wang2022debiasing}, we assume the attacker has a dataset that contains users' ratings of items. This dataset can be obtained via generative methods or crawled from the internet~\cite{zhu2023membership}, and it is used for generating item features using matrix factorization~\cite{koren2009matrix}. Unlike the existing works in~\cite{zhang2021membership,wang2022debiasing}, we do not assume the attacker has a shadow dataset that comes from the same distribution as the training dataset of the recommender system. In addition, we do not assume the attacker has knowledge of the model architecture of the recommender system. In contrast, the availability of such knowledge is a key factor for the success of MIAs in existing works~\cite{zhang2021membership,wang2022debiasing} following shadow training.

\subsection{Shadow-based MIAs and Their Limitations}\label{sec:shadow-training}
\paragraph{Notations.} Let $\bm{M}^{p \times q}$ be a user-item matrix that contains $p$ users' ratings of $q$ items. Using the matrix factorization~\cite{koren2009matrix} technique, we can divide $\bm{M}^{p \times q}$ into the product of two lower dimensional matrices:
\begin{equation}
    \hat{\bm{M}}^{p \times q}=\bm{H}^{p \times l}\cdot\bm{W}^{l \times q},
\end{equation}
by minimizing the following loss function:
\begin{equation}
    \min || {\bm{M}}^{p \times q} - \hat{\bm{M}}^{p \times q} ||_2,
\end{equation}
where $||\cdot||_2$ is the $l$-2 norm. $\bm{H}$ contains the user's latent factors, $\bm{W}^{\textrm{T}}$ contains item's latent factors. We denote $\bm{W}^{\textrm{T}}={(\bm{w}_1;\cdots;\bm{w}_q})$, where each $\bm{w}_i$ is a $l$-dimensional vector and represents the feature of an item. Let $\bm{x}=[x_1,\cdots,x_m]$ be $m$ historical interactions of a user, where each $x_i$ is an item. A recommender system is a function $f(\cdot)$ that takes as input the historial interactions of a user and outputs $n$ recommended items $\bm{Y}=f(\bm{x})$ to the user. Specifically, $\bm{Y}=[y_1,\cdots,y_n]$ is a $n$-dimensional vector and each $y_i$ represents a recommended item.

\noindent\textbf{Existing Shadow-based Membership Inference.} Shadow-based MIAs aim to train a binary classifier $h(\cdot)$ that takes as an input a user's feature vector $\bm{v}$ and outputs 0 or 1:

\begin{equation}
    h: \bm{V} \rightarrow \{0,1\},
\end{equation}
where $\bm{V}$ represents users' feature vector space and $\bm{v} \in \bm{V}$, 0 represents that the attack classifier predicts the user as a non-member, and 1 as a member. To train the binary attack classifier, the attacker requires to obtain the features of member users and non-member users. 

Existing shadow-based MIAs assume the attacker can have a shadow dataset $\mathcal{D}_s$ that comes from the same distribution of the training data of the target recommender system. In addition, the attacker is assumed to know the model architecture $\mathcal{A}$ of the target recommender system. There are three steps in the shadow-based MIAs: \textit{i)} The attacker splits $\mathcal{D}_s$ into two disjoint datasets: $\mathcal{D}_s^{\textrm{train}}$ and $\mathcal{D}_s^{\textrm{test}}$. Following the same model architecture of $\mathcal{A}$, the attacker trains a shadow model $f_s(\cdot)$ on $\mathcal{D}_s^{\textrm{train}}$ to mimic the behavior of the target model; \textit{ii)} After the training of $f_s(\cdot)$, for each user's historical interactions $\bm{x} \in \mathcal{D}_s^{\textrm{train}}$, the attacker queries the shadow model and records the corresponding recommended items $\bm{Y}_{\textrm{train}}=[y_1,\cdots,y_n]$. Since the attacker has $\bm{W}^{\textrm{T}}$, the attacker can obtain a feature vector as:
\begin{equation}
    \bm{v}=\frac{1}{m}\sum_{m=1}^{m}\bm{w}_{x_m}-\frac{1}{n}\sum_{n=1}^{n}\bm{w}_{y_n},
\end{equation}
where $\bm{v}$ is a $l$-dimensional vector and $\bm{w}_i$ is the corresponding feature vector of the item $i$. The attacker considers such a vector as the feature vector of a user since it encodes information on both recommendations and historical interactions. Based on $\bm{x} \in \mathcal{D}_s^{\textrm{train}}$ and $\bm{x} \in \mathcal{D}_s^{\textrm{test}}$, the attacker can obtain feature vectors of member and non-member users by querying the shadow model; \textit{iii)} Since the attacker has collected features of users in the previous step, the attacker now can train a binary classifier using standard machine learning training procedures. After training, the binary classifier works as an attack model for predicting the membership status of a target user of the target recommender system.

\begin{figure*}[t]
\centering
\includegraphics[width=1\linewidth]{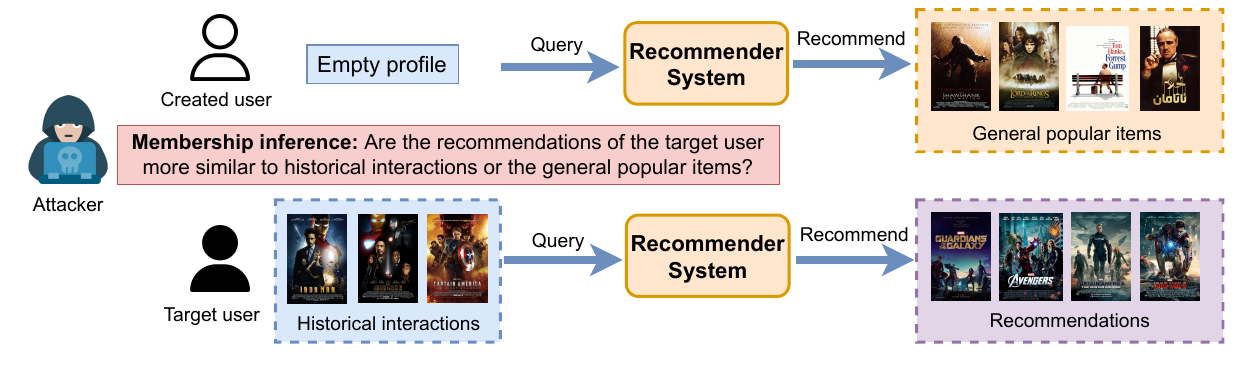}
 \caption{An overview of shadow-free MIAs. The attacker creates a user with an empty profile to obtain the general popular items of the recommender system. For a target user, the attacker examines whether the recommendations of the target user are more similar to his historical interactions or the general popular items to determine the membership status of the target user.}
\label{fig:mia}
\end{figure*}

\noindent\textbf{Limitations of Existing Attacks.} There are two limitations to existing shadow-based MIAs. First, shadow-based MIAs can not work effectively when the target dataset is not available or when the model architecture of the target recommender system is unknown. This is because, under this scenario, the shadow model cannot mimic the behavior of the target model well. Second, constructing the attack model is relatively computationally expensive because the attacker is required to train a shadow model and an attack model. When both of the models are complex, following the same architecture, the attacker requires large computational resources for training the shadow model. In Section~\ref{sec:effectiveness}, we demonstrate that in some cases where the target model is a deep learning-based recommender system, it takes a long time to train the attack model. To solve the two limitations, we propose shadow-free MIAs: an attacker is able to efficiently and effectively conduct membership inference without the training of a shadow model and the requirements of the target dataset or knowledge of the target recommender system.

\subsection{Shadow-free Membership Inference}\label{sec:shadow-free}
\paragraph{Key Intuition.} Figure~\ref{fig:mia} shows an overview of shadow-free MIAs. The key of the attack is to compare the recommendations of the target user to his historical interactions and the general popular items. A key observation is leveraged to determine the membership status of a user: A member user's recommendations provided by the recommender system are similar to his interactions because such interactions are used by the recommender system to find relevant items. As depicted in Figure~\ref{fig:mia}, if a user likes Marvel movies and his data was used by the recommender system, it is highly likely to recommend other Marvel movies to the user. This makes the recommendations of a member user much more similar to the interactions than the general popular items recommended by the recommender system, e.g., general high-rated movies depicted in Figure~\ref{fig:mia}. Thus, the attacker can infer a user as a member if the recommendations are more similar to the interactions than the general popular items, and infer the user as a non-member otherwise.

Formally, there are three steps for conducting shadow-free MIAs, described as follows:

\paragraph{i) Creating an Empty User.} To obtain the general popular items of the recommender system, the attacker first creates a user account with no interactions with the recommender system. This is not difficult to achieve in practice, e.g., the attacker can easily register a new account in IMDB. Because the attacker has black-box access to the recommender system, the attacker can obtain $n$ popular items using the newly created user account. We define these popular items as:
\begin{equation}
    \bm{Y}_p=[y_1,\cdots,y_n].
\end{equation}
Using the item's latent matrix $\bm{W}^{\textrm{T}}$ and each $y_i$ in $\bm{Y}_p$, the attacker can obtain a feature vector of $\bm{Y}_p$:
\begin{equation}
    \bm{v}_p=\frac{1}{n}\sum_{n=1}^{n} \bm{w}_{y_n}.
\end{equation}

\paragraph{ii) Query the Recommender System.} For a target user with historical interactions $\bm{X}=[x_1,\cdots,x_m]$, the attacker queries the target model and obtains $n$ recommended items. We define these recommendations as:
\begin{equation}
    \bm{Y}_t=[y_1,\cdots,y_n].
\end{equation}
Using the item's latent matrix $\bm{W}^{\textrm{T}}$, each $x_i$ in $\bm{X}$, and each $y_i$ in $\bm{Y}_t$, the attacker can obtain a feature vector of $\bm{X}$ and $\bm{Y}_t$, respectively:
\begin{equation}
    \bm{v}_x=\frac{1}{m}\sum_{m=1}^{m} \bm{w}_{x_m},
\end{equation}
\begin{equation}
    \bm{v}_t=\frac{1}{n}\sum_{n=1}^{n} \bm{w}_{y_n}.
\end{equation}

\paragraph{iii) Infer the Membership Privacy.} To determine the membership status of the target user, the attacker needs to determine whether the recommendations are more similar to the interactions or the general popular items. This can be done by calculating the corresponding similarities and comparing them. Given the feature vectors $\bm{v}_p$, $\bm{v}_x$, and $\bm{v}_t$, the attacker calculates two distances:
\begin{equation}
    \alpha_1=||\bm{v}_p - \bm{v}_t||_2,
\end{equation}
\begin{equation}
    \alpha_2=||\bm{v}_x - \bm{v}_t||_2.
\end{equation}

In the context of recommender systems, $\alpha_1$ represents how close the recommendations are to the general popular items, while $\alpha_2$ represents how similar the recommendations are to the interactions of the target user. In general, a larger $\alpha$ represents a smaller similarity. If $\alpha_1>\alpha_2$, indicating that the recommendations are more similar to the interactions of the target user, the attacker considers the user as a member. Otherwise, the attacker considers the user as a non-member. Formally, the attack $\mathcal{M}(\cdot,\cdot)$ is defined as follows:
\begin{equation}
\mathcal{M}(\alpha_1,\alpha_2) = \begin{cases}
1 \quad {\textrm{if} \; \alpha_1>\alpha_2, }\\
0 \quad {\textrm{otherwise}}.
\end{cases}
\end{equation}

\begin{table*}

\centering
 
  \resizebox{\linewidth}{!}{\begin{tabular}{cccccccccccccccc}
\toprule
& \multicolumn{15}{c}{\textbf{Target recommender systems}} \\
\cmidrule{2-16}
\multicolumn{1}{c}{\multirow{2}{*}{\textbf{Attacks}}} & \multicolumn{3}{c}{ICF} & \multicolumn{3}{c}{NCF} & \multicolumn{3}{c}{BERT4Rec} & \multicolumn{3}{c}{Caser} & \multicolumn{3}{c}{GRU4Rec}
\\
\cmidrule{2-16}
\multicolumn{1}{c}{} &Accuracy & TPR & FPR &Accuracy  & TPR & FPR &Accuracy  & TPR & FPR &Accuracy  & TPR & FPR &Accuracy & TPR & FPR   \\

\cmidrule(r){2-4}
\cmidrule(r){5-7}
\cmidrule(r){8-10}
\cmidrule(r){11-13}
\cmidrule(r){14-16}

\multicolumn{1}{c}{SF-MIAs (Ours)} & \textbf{0.793} &0.611 & \textbf{0.025} & \textbf{0.968} &0.960 &\textbf{0.025} &  0.808 &0.675 & \textbf{0.059} &  \textbf{0.986} & 0.997 &\textbf{0.025} &  \textbf{0.983} &0.999 &0.033 \\
\cmidrule(r){2-4}
\cmidrule(r){5-7}
\cmidrule(r){8-10}
\cmidrule(r){11-13}
\cmidrule(r){14-16}

\multicolumn{1}{c}{\multirow{2}{*}{ST-MIA}} &BT 0.500 &0.999 &1.000 & BT 0.499 &0.997 &1.000 & GA 0.403  &0.451 &0.645 & BA 0.500&\textbf{1.000} &1.000 &CA 0.503 &\textbf{1.000} &0.995 \\
\multicolumn{1}{c}{} &NT 0.500 &\textbf{1.000} &1.000 & IT 0.502 &\textbf{1.000} &0.996 & CA 0.493 &0.982 &0.997 & GA 0.670 &0.999 &0.658 &BA 0.500
 &\textbf{1.000} &1.000\\
\cmidrule(r){2-4}
\cmidrule(r){5-7}
\cmidrule(r){8-10}
\cmidrule(r){11-13}
\cmidrule(r){14-16}

 \multicolumn{1}{c}{\multirow{2}{*}{DL-MIA}} &BT 0.582 &\textbf{1.000} &0.839 & BT 0.502 &\textbf{1.000} &1.000 & GA 0.340 &0.001 &0.199 & BA 0.930 &0.860 &0.124 & CA 0.950  & \textbf{1.000} & \textbf{0.032} \\
\multicolumn{1}{c}{} &NT 0.504 & \textbf{1.000} & 0.991 & IT 0.502 &\textbf{1.000} &1.000 & CA \textbf{0.868} &\textbf{1.000} &0.266 & GA 0.984 & \textbf{1.000}& 0.032 & BA 0.884 
 &\textbf{1.000} &0.233\\
\bottomrule 
\end{tabular}
}
\caption{Attack accuracy, TPR, and FPR of the shadow-free MIAs (SF-MIAs) and the two attack baselines across five recommender systems on the MovieLens-1M dataset.}
  \label{tab:movielens}

\end{table*}

Essentially, our shadow-free MIA is metric-based MIA, analogous to metric-based MIAs in the context of classification models~\cite{yeom2018privacy,song2021systematic} where prediction vectors of data records are used for calculating metrics and the calculated metrics are then compared with a preset threshold to determine the membership status of data records. However, different from these metric-based MIAs, the metric (i.e., comparing $\alpha_1$ with $\alpha_2$) in our attacks is self-adaptive as it is calculated in a user-specific manner. Thus, in our attacks, preset threshold is uncessary. The self-adaptiveness is a big advantage of our method compared to other metric-based approaches where determining the threshold value is a non-trivial job for an attacker. The technical innovations of shadow-free MIAs bring substantial benefits including a simplified methodology design, wide applicability with a practical black-box assumption, improved attack accuracy, and low computation cost.

\section{Experiments}\label{sec:experiments}

\subsection{Experimental Setup}

\begin{table*}
\small
\centering
  
  \resizebox{\linewidth}{!}{\begin{tabular}{cccccccccccccccc}
\toprule
& \multicolumn{15}{c}{\textbf{Target recommender systems}} \\
\cmidrule{2-16}
\multicolumn{1}{c}{\multirow{2}{*}{\textbf{Attacks}}} & \multicolumn{3}{c}{ICF} & \multicolumn{3}{c}{NCF} & \multicolumn{3}{c}{BERT4Rec} & \multicolumn{3}{c}{Caser} & \multicolumn{3}{c}{GRU4Rec}
\\
\cmidrule{2-16}
\multicolumn{1}{c}{} &Accuracy  & TPR & FPR &Accuracy  & TPR & FPR &Accuracy  & TPR & FPR &Accuracy  & TPR & FPR &Accuracy  & TPR & FPR   \\

\cmidrule(r){2-4}
\cmidrule(r){5-7}
\cmidrule(r){8-10}
\cmidrule(r){11-13}
\cmidrule(r){14-16}

\multicolumn{1}{c}{SF-MIAs (Ours)} & \textbf{0.815} &0.630 &\textbf{0.000} & \textbf{0.620} &0.241 &\textbf{0.000} & \textbf{0.693} &0.386 &\textbf{0.000} & 0.727&0.453 &\textbf{0.000} & \textbf{0.923}  &0.845 &\textbf{0.000} \\
\cmidrule(r){2-4}
\cmidrule(r){5-7}
\cmidrule(r){8-10}
\cmidrule(r){11-13}
\cmidrule(r){14-16}

\multicolumn{1}{c}{\multirow{2}{*}{ST-MIA}} &BM 0.520 &0.517 &0.478 & IM 0.571 &0.602 &0.461 & CM 0.543  &0.117 &0.031 & BM 0.584 &0.472 &0.304 &CM 0.890 &0.796 &0.016 \\
\multicolumn{1}{c}{} &BT 0.330 &\textbf{0.636} &0.976 & BT 0.414 &0.804 &0.976 & GM 0.553 &0.153 &0.048 & GM 0.736 &0.510 &0.038 &BM 0.673
 &0.633 &0.287\\
\cmidrule(r){2-4}
\cmidrule(r){5-7}
\cmidrule(r){8-10}
\cmidrule(r){11-13}
\cmidrule(r){14-16}

\multicolumn{1}{c}{\multirow{2}{*}{DL-MIA}} &BM 0.468 &0.280 &0.343 & IM 0.503 &\textbf{0.938} &0.933 & CM 0.509  &0.922 &0.904 & BM 0.519 &\textbf{0.991} &0.954 &CM 0.503 &\textbf{0.913} &0.894 \\
\multicolumn{1}{c}{} &BT 0.498 &0.007 &0.010 & BT 0.505 &0.014 &0.005 & GM 0.572 &\textbf{0.958} &0.816 & GM \textbf{0.916} &0.951 &0.120 &BM 0.346
 &0.314 &0.623 \\

\bottomrule 
\end{tabular}
}

\caption{Attack accuracy, TPR, and FPR of the shadow-free MIAs (SF-MIAs) and the two attack baselines across five recommender systems on the Amazon Beauty dataset.}
  \label{tab:beauty}
\end{table*}

We conduct extensive experiments on three benchmark datasets across five different recommender systems, which include traditional recommender system and advanced deep learning based ones. Due to page limits, the detailed description of datasets, dataset partition, main parameter settings in recommender systems, detailed introduction of baselines, baseline attack setting descriptions, and facilities utilized for experiments are available in Appendix~\ref{appendix::dataset}\footnote{Please refer to the version of this paper with Appendix in arXiv.}.

\paragraph{Datasets.} In the experiments, three benchmark datasets are leveraged: MovieLens-1M~\cite{harper2015movielens}, Amazon Beauty~\cite{mcauley2015image}, and Ta-feng \footnote{https://www.kaggle.com/datasets/chiranjivdas09/ta-feng-grocery-dataset}. All these datasets are benchmark datasets for evaluating the performance of recommender systems.

\paragraph{Recommender Systems.} We select five representative recommender systems to comprehensively evaluate our proposed attacks. These recommender systems including the traditional recommender system of the Item-based Collaborative Filtering (ICF)~\cite{sarwar2001item}, as well as the advanced deep learning based ones of the Neural Collaborative Filtering (NCF)~\cite{he2017neural}, BERT4Rec~\cite{sun2019bert4rec}, Caser~\cite{tang2018personalized}, and GRU4Rec~\cite{hidasi2015session}. Following previous works~\cite{zhang2021membership,wang2022debiasing}, personalized recommendation lists are generated for existing users via  recommendation algorithms of the recommender systems. For new users, due to a lack of their data, recommender systems recommend the most popular items to them.

\paragraph{Baselines.} We compare the proposed shadow-free MIAs (SF-MIAs) with two state-of-the-art attack methods: shadow-based MIAs (ST-MIAs)~\cite{zhang2021membership} and Debiasing learning for MIAs (DL-MIAs)~\cite{wang2022debiasing}. The two baselines in our experiments are with the same black-box assumption as our approach.

\paragraph{Evaluation Metrics.}  We evaluate MIAs from two perspectives: effectiveness and efficiency. In terms of effectiveness, since membership inference is a binary classification problem, we use accuracy to evaluate the attack performance, which is one of the most widely used metrics in existing studies of MIAs~\cite{hu2022membership}. In addition, as suggested by~\cite{carlini2022membership} that a powerful and reliable MIA should have a high true positive rate (TPR) at a low false positive rate (FPR), we report TPR and FPR of our attacks and the two baselines. An attack is considered to be reliable if it can achieve a high true positive rate at a very low false positive rate. In terms of efficiency, we record the overall computational time of the attack model of different approaches. An attack that has a short overall computational time is considered as efficient.

\paragraph{Attack Settings.} For baseline methods, we run experiments against combinations of different shadow models and shadow datasets to achieve a comprehensive and fair evaluation.

\subsection{Efficacy of The Proposed Attack}\label{sec:effectiveness}

\begin{table*}
\small
\centering

  \resizebox{\linewidth}{!}{\begin{tabular}{cccccccccccccccc}
\toprule
& \multicolumn{15}{c}{\textbf{Target recommender systems}} \\

\cmidrule{2-16}

\multicolumn{1}{c}{\multirow{2}{*}{\textbf{Attacks}}} & \multicolumn{3}{c}{ICF} & \multicolumn{3}{c}{NCF} & \multicolumn{3}{c}{BERT4Rec} & \multicolumn{3}{c}{Caser} & \multicolumn{3}{c}{GRU4Rec}\\

\cmidrule{2-16}
\multicolumn{1}{c}{} &Accuracy  & TPR & FPR &Accuracy  & TPR & FPR &Accuracy  & TPR & FPR &Accuracy  & TPR & FPR &Accuracy  & TPR & FPR   \\

\cmidrule(r){2-4}
\cmidrule(r){5-7}
\cmidrule(r){8-10}
\cmidrule(r){11-13}
\cmidrule(r){14-16}

\multicolumn{1}{c}{SF-MIAs (Ours)}  & \textbf{0.981} & 0.961& \textbf{0.000} & 0.856 & 0.713 & \textbf{0.000} & \textbf{0.645} & 0.210 & \textbf{0.000} & \textbf{0.977} & 0.954 & \textbf{0.000} &  \textbf{0.991}& 0.983 & \textbf{0.000} \\
\cmidrule(r){2-4}
\cmidrule(r){5-7}
\cmidrule(r){8-10}
\cmidrule(r){11-13}
\cmidrule(r){14-16}

\multicolumn{1}{c}{\multirow{2}{*}{ST-MIA}} &CA 0.950 &0.998 &0.098 & GA 0.875 &0.916 & 0.165 & CA 0.579  &0.244 &0.087 & GA 0.897&\textbf{0.959} &0.165 &BA 0.572 &\textbf{1.000} &0.855 \\
\multicolumn{1}{c}{} &BM 0.897 & 0.829 & 0.036 & BM 0.673 & 0.381&0.036 & GM 0.495 &0.002 &0.013 & NM 0.522 &0.947 &0.902 &BM 0.922 &0.875 &0.031\\
\cmidrule(r){2-4}
\cmidrule(r){5-7}
\cmidrule(r){8-10}
\cmidrule(r){11-13}
\cmidrule(r){14-16}

\multicolumn{1}{c}{\multirow{2}{*}{DL-MIA}} &CA 0.500 &\textbf{1.000} &1.000 & GA 0.939 &0.991 & 0.113 & CA 0.499  &\textbf{0.999} &0.999 & GA 0.889 &0.895 &0.117 &BA 0.939 &0.904 &0.105 \\
\multicolumn{1}{c}{} &BM 0.682 & 0.724 & 0.360 & BM \textbf{0.947} & \textbf{0.993}&0.099 & GM 0.449 &0.005 &0.006 & NM 0.703 &0.682 &0.277 &BM 0.696 &0.679 &0.288 \\
 
\bottomrule 
\end{tabular}
}

\caption{Attack accuracy, TPR, and FPR of the shadow-free MIAs (SF-MIAs) and the two attack baselines across five recommender systems on the Ta-feng dataset.}
  \label{tab:taFeng}
  
\end{table*}

\begin{table}
\small

\centering
\resizebox{\linewidth}{!}{\begin{tabular}{lccc}
\toprule 
Method & Shadow-free MIAs & ST-MIA & DL-MIA\\
\midrule
Time cost (avg) & 3.7s & 128.8s ($\approx 35\times$) & 9,760s ($\approx 2,637\times$) \\
\bottomrule
\end{tabular}}
\caption{The computational cost of shadow-free MIAs and baseline attacks, measured by seconds. ($T\times$) represents $T$ times faster of the shadow-free MIAs than the baselines.}
\label{tab:time}
\end{table}

\paragraph{Attack Effectiveness.} 
Table~\ref{tab:movielens}, Table~\ref{tab:beauty}, and Table~\ref{tab:taFeng} show attack accuracy, TPR, and FPR of our attack and the two baseline attacks. Based on the experimental results, we can observe that: (\textit{i}) In general, our proposed shadow-free MIAs have very high attack accuracy. The accuracy scores of our method across all attack settings are above 60\%. In addition, the accuracy of our proposed attack is above 80\% across 70\% of attack settings, demonstrating the vulnerabilities of recommender systems in most settings. In some cases, our attacks achieve perfect performance with an accuracy close to 100\%. For example, in Table~\ref{tab:movielens}, the highest accuracy of our method can be achieved at 98.6\% when the Caser recommender system is trained on MovieLens-1M; (\textit{ii}) Our shadow-free MIAs consistently outperform the baselines in all experimental settings in Table~\ref{tab:movielens} with respect to accuracy. For Table~\ref{tab:beauty} and Table~\ref{tab:taFeng}, the accuracy of our attacks is higher than the two baselines in almost 80\% experimental settings. In addition, in some cases, our attack can achieve near-perfect performance, while the baselines have performance close to random guess. For example, in Table~\ref{tab:movielens}, when NCF is trained on MovieLens-1M, the accuracy of our attacks is 96.8\%, while the attack accuracy of baselines is nearly close to 50\%. (\textit{iii}) MIAs based on shadow training (i.e., the two baselines) have severe limitations when the attacker does not know the training data distribution and the model architecture of the target recommender system. For example, in Table~\ref{tab:movielens}, the accuracy of ST-MIA is close to 50\% (random guess) in almost 80\% experimental settings. For DL-MIAs, although attack accuracy is improved in most cases, they still perform worse than our attacks. 

\noindent\textbf{Takeaway 1.~}
\textit{The proposed shadow-free MIAs are more effective than the two baselines. In most cases, the proposed attacks achieve an attack accuracy higher than 0.8, while the baseline attacks can only achieve performance close to random guess with an accuracy of around 0.5.}

\paragraph{Attack Reliability.} 
As suggested by~\cite{carlini2022membership}, a powerful and reliable MIA should have a high TPR at a low FPR. From the experimental results in Table~\ref{tab:movielens}, Table~\ref{tab:beauty}, and Table~\ref{tab:taFeng}, our attack can achieve high TPRs at low FPRs close to 0. For instance, in Table~\ref{tab:movielens}, when GRU4Rec is trained on MovieLens-1M,  our attack achieves a TPR of 99.9\% with an FPR of 3.3\%. In Table~\ref{tab:beauty}, when GRU4Rec is trained on Amazon Beauty, our attack achieves a TPR of 84.5\% with an FPR of 0, demonstrating that no non-member users are mistakenly predicted as member users. In Table~\ref{tab:taFeng}, when the target recommender system is ICF, NCF, Caser, and GRU4Rec, the TPR of our attack is 96.1\%, 71.3\%, 95.4\%, and 98.3\% with FPRs of 0. In contrast, the FPR values tend to be high for baselines. For instance, in Table~\ref{tab:beauty}, when target recommender NCF is trained on the target dataset Amazon Beauty and shadow recommender BERT4Rec is trained on Ta-feng, ST-MIA attack achieves a TPR of 80.4\% with an FPR of 97.6\%. When the target recommender model of NCF is trained on the target dataset Amazon Beauty, the shadow model using ICF trained on MovieLens-1M, DL-MIA attack achieves a TPR of 93.8\% with an FPR of 93.3\%. The experimental results demonstrate that our proposed SF-MIA is more powerful and reliable.

\noindent \textbf{Takeaway 2.~}
\textit{The shadow-free MIAs are more reliable than baseline attacks. In most cases, the proposed attacks achieve high TPRs with low FPRs close to 0, while the baselines have high FPRs in most cases.}

\paragraph{Attack Efficiency.} To compare the efficiency of attack methods, we record the overall computational time cost of different attack methods. For each method, we select five experimental settings to run, and the average time cost is calculated. For instance, we select ``(BERT., Ama.)'', ``(BERT., Mov.)'', ``(BERT., Ta.)'', ``(Cas., Ama.)'', and ``(Cas., Mov.)'' settings to calculate the time cost for SF-MIA. For ST-MIA and DL-MIA, we select settings ``IM'' from Table~\ref{tab:beauty}, settings ``NT'', ``IT'' from Table~\ref{tab:movielens}, and settings ``NM'' and ``BA'' from Table~\ref{tab:taFeng} as experimental settings to calculate time cost. As we can see in Table~\ref{tab:time}, our attack method takes only about 3.7 seconds to complete the attack. In contrast, ST-MIA takes almost 128.8 seconds to implement the attack, and DL-MIA is rather expensive in terms of time cost, with around 9,760 seconds. This validates that our attacks are much more efficient compared with baseline attacks.

\noindent \textbf{Takeaway 3.~}
\textit{The shadow-free MIAs are more efficient than baseline attacks.}

\subsection{Why Shadow-free MIAs Work}

In shadow-free MIAs, the attacker needs to calculate and compare two distances for a target user to determine the membership status. To understand why the proposed attacks work, we provide a visualization of $\alpha_1 - \alpha_2$ (see Section~\ref{sec:shadow-free} for the definition of $\alpha$) for member users and non-member users in Figure~\ref{fig:violin}. In our proposed attacks, 0 essentially is the threshold of $\alpha_1-\alpha_2$ to determine whether a target user is a member or a non-member. We select four attack settings and calculate the difference value between $\alpha_1$ and $\alpha_2$, i.e., $\alpha_1 - \alpha_2$. From the visualization, we can see that the distributions of member and non-member users are very different, explaining why our proposed attacks are highly effective. Specifically, we can observe that in the setting where GRU4Rec is trained on Amazon Beauty, all non-members' data is less than 0, which means that all non-members are correctly classified, which explains that in Table~\ref{tab:beauty}, the FPR of the setting is equal to 0. In addition, the dotted line almost overlapped with the mid-line of data distribution of member data in the setting where ICF is trained on MovieLens-1M, which shows that shadow-free MIA can predict about 50\% of members correctly for this setting. For the settings where Caser is trained on MovieLens-1M and GRU4Rec is trained on MovieLens-1M, a high portion of members are correctly classified, and less than half of non-members are predicted as members. Due to page limits, we provide the corresponding experimental quantity results in Appendix~\ref{appendix::results}.

\subsection{Ablation Study}

We analyse how different parameters of the number of recommends and the length of items' latent feature vectors can influence the attack performance of the proposed attack. Due to page limits, we present the findings from the experiments, while providing the quantity results in Appendix~\ref{appendix::results}.

\paragraph{The Number of Recommendations $n$.} We use GRU4Rec trained on MovieLens-1M to study how the number of recommendations can influence the attack performance. We vary the number of recommendations from 10 to 100. Experimental results show that increasing the number of recommendations slightly decreases the attack accuracy. This might be because more recommendations can add general popular items to member users' recommendations, making it more difficult to distinguish member users from non-members. Nonetheless, the attack accuracy remains above 97.5\% in all cases, demonstrating the high effectiveness of the attack.

\paragraph{The Length of Vectors $l$.} We use GRU4Rec trained on Ta-feng to study how the length of the item feature vector can influence the attack performance. We vary the number of length of item feature vectors from 10 to 100.  Similar to the study of the number of recommendations, the attack performance is stable at different lengths, and the attack accuracy is all above 98\%, demonstrating the high effectiveness of the attack.

\noindent \textbf{Takeaway 4.~}
\textit{The shadow-free MIAs are stable and effective under different attack settings. The number of recommendations and the length of the item feature vector only slightly influence the attack performance.}

\section{Discussion}
As demonstrated in Section~\ref{sec:experiments}, shadow-free MIAs can infer the membership privacy of a user efficiently and effectively. As our attacks only need black-box access to the model, the experimental results in this paper shed light on the vulnerabilities of recommender systems in leaking the membership privacy of their member users. Two possible defenses can be considered to mitigate MIAs. The first one is leveraging differential privacy~\cite{dwork2006calibrating}, the most widely used privacy mechanism, to train a differentially private recommender system that should not remember the details of a specific user. For deep learning-based recommender systems, DP-SGD~\cite{abadi2016deep} can be leveraged during the training process. Another potential defense is to add randomness to the recommendations for non-members. For example, except for recommending general popular items to non-member users, the recommender system can randomly select some items from the whole recommendation lists and add them to the final recommendations. This can make non-member users' recommendations similar to personalized recommendations of member users, which may make the attacker mistakenly predict non-members as members, reducing the reliability of the MIAs.

\paragraph{Limitations.}
One limitation of our method is that the target recommender system leverages the popularity-based recommendation strategy to handle the cold-start problem for new users. While this simple yet effective strategy is still a mainstream solution~\cite{sedhain2014social}, more strategies employing the information from other sources (e.g., users' social data~\cite{sedhain2017low}) to approximate user preference have also been proposed. We believe that the attack principle in our method can also be applied in such a setting, but some tweaks might be necessary. We leave this part of research in our future work.

\section{Conclusion}
In this paper, we proposed shadow-free MIAs that can effectively and efficiently infer the membership privacy of a user in recommender systems. Compared to existing works that require training a shadow model on a shadow dataset, our attack requires only black-box access to the target recommender system. We conduct extensive experiments on three benchmark datasets across five recommender systems under different attack settings. The experimental results validate that our attack can efficiently and effectively achieve high attack accuracy at a low false positive rate, which is far better than baseline attacks. The findings in this paper shed light on the vulnerability of recommender systems, emphasizing the importance of comprehensively evaluating the privacy
risks of recommender systems. Two possible defenses that can mitigate the membership privacy leakage of recommender systems are discussed. We leave the evaluation of the effectiveness of such defenses in future works.

\clearpage

\section*{Acknowledgments}
Dr Xuyun Zhang is the recipient of an ARC DECRA (project No. DE210101458) funded by the Australian Government. Professor Lianyong Qi is supported by Natural Science Foundation of Shandong Province (No. ZR2023MF007).

\bibliographystyle{named}
\bibliography{ijcai24}

\begin{thebibliography}{}

\bibitem[\protect\citeauthoryear{Abadi \bgroup \em et al.\egroup }{2016}]{abadi2016deep}
Martin Abadi, Andy Chu, Ian Goodfellow, H~Brendan McMahan, Ilya Mironov, Kunal Talwar, and Li~Zhang.
\newblock Deep learning with differential privacy.
\newblock In {\em CCS}, pages 308--318, 2016.

\bibitem[\protect\citeauthoryear{Burke}{2002}]{burke2002hybrid}
Robin Burke.
\newblock Hybrid recommender systems: Survey and experiments.
\newblock {\em User Modeling and User-adapted Interaction}, 12:331--370, 2002.

\bibitem[\protect\citeauthoryear{Carlini \bgroup \em et al.\egroup }{2019}]{carlini2019secret}
Nicholas Carlini, Chang Liu, {\'U}lfar Erlingsson, Jernej Kos, and Dawn Song.
\newblock The secret sharer: Evaluating and testing unintended memorization in neural networks.
\newblock In {\em USENIX Security}, pages 267--284, 2019.

\bibitem[\protect\citeauthoryear{Carlini \bgroup \em et al.\egroup }{2021}]{carlini2021extracting}
Nicholas Carlini, Florian Tramer, Eric Wallace, Matthew Jagielski, Ariel Herbert-Voss, Katherine Lee, Adam Roberts, Tom Brown, Dawn Song, Ulfar Erlingsson, et~al.
\newblock Extracting training data from large language models.
\newblock In {\em USENIX Security}, pages 2633--2650, 2021.

\bibitem[\protect\citeauthoryear{Carlini \bgroup \em et al.\egroup }{2022}]{carlini2022membership}
Nicholas Carlini, Steve Chien, Milad Nasr, Shuang Song, Andreas Terzis, and Florian Tramer.
\newblock Membership inference attacks from first principles.
\newblock In {\em S\&P}, pages 1897--1914. IEEE, 2022.

\bibitem[\protect\citeauthoryear{Chen \bgroup \em et al.\egroup }{2017}]{chen2017attentive}
Jingyuan Chen, Hanwang Zhang, Xiangnan He, Liqiang Nie, Wei Liu, and Tat-Seng Chua.
\newblock Attentive collaborative filtering: Multimedia recommendation with item-and component-level attention.
\newblock In {\em SIGIR}, pages 335--344, 2017.

\bibitem[\protect\citeauthoryear{Ding \bgroup \em et al.\egroup }{2018}]{ding2018improving}
Jingtao Ding, Guanghui Yu, Xiangnan He, Yuhan Quan, Yong Li, Tat-Seng Chua, Depeng Jin, and Jiajie Yu.
\newblock Improving implicit recommender systems with view data.
\newblock In {\em IJCAI}, pages 3343--3349, 2018.

\bibitem[\protect\citeauthoryear{Dwork \bgroup \em et al.\egroup }{2006}]{dwork2006calibrating}
Cynthia Dwork, Frank McSherry, Kobbi Nissim, and Adam Smith.
\newblock Calibrating noise to sensitivity in private data analysis.
\newblock In {\em Theory of Cryptography: Theory of Cryptography Conference (TCC)}, pages 265--284. Springer, 2006.

\bibitem[\protect\citeauthoryear{Fan \bgroup \em et al.\egroup }{2019}]{fan2019graph}
Wenqi Fan, Yao Ma, Qing Li, Yuan He, Eric Zhao, Jiliang Tang, and Dawei Yin.
\newblock Graph neural networks for social recommendation.
\newblock In {\em WWW}, pages 417--426, 2019.

\bibitem[\protect\citeauthoryear{Harper and Konstan}{2015}]{harper2015movielens}
F~Maxwell Harper and Joseph~A Konstan.
\newblock The movielens datasets: History and context.
\newblock {\em Acm Transactions on Interactive Intelligent Systems (TIIS)}, 5(4):1--19, 2015.

\bibitem[\protect\citeauthoryear{Hayes \bgroup \em et al.\egroup }{2017}]{hayes2017logan}
Jamie Hayes, Luca Melis, George Danezis, and Emiliano De~Cristofaro.
\newblock Logan: Membership inference attacks against generative models.
\newblock {\em arXiv preprint arXiv:1705.07663}, 2017.

\bibitem[\protect\citeauthoryear{He \bgroup \em et al.\egroup }{2016}]{he2016fast}
Xiangnan He, Hanwang Zhang, Min-Yen Kan, and Tat-Seng Chua.
\newblock Fast matrix factorization for online recommendation with implicit feedback.
\newblock In {\em SIGIR}, pages 549--558, 2016.

\bibitem[\protect\citeauthoryear{He \bgroup \em et al.\egroup }{2017}]{he2017neural}
Xiangnan He, Lizi Liao, Hanwang Zhang, Liqiang Nie, Xia Hu, and Tat-Seng Chua.
\newblock Neural collaborative filtering.
\newblock In {\em WWW}, pages 173--182, 2017.

\bibitem[\protect\citeauthoryear{He \bgroup \em et al.\egroup }{2020}]{he2020segmentations}
Yang He, Shadi Rahimian, Bernt Schiele, and Mario Fritz.
\newblock Segmentations-leak: Membership inference attacks and defenses in semantic image segmentation.
\newblock In {\em Computer Vision--ECCV}, pages 519--535. Springer, 2020.

\bibitem[\protect\citeauthoryear{He \bgroup \em et al.\egroup }{2021}]{he2021node}
Xinlei He, Rui Wen, Yixin Wu, Michael Backes, Yun Shen, and Yang Zhang.
\newblock Node-level membership inference attacks against graph neural networks.
\newblock {\em arXiv preprint arXiv:2102.05429}, 2021.

\bibitem[\protect\citeauthoryear{Hidasi \bgroup \em et al.\egroup }{2015}]{hidasi2015session}
Bal{\'a}zs Hidasi, Alexandros Karatzoglou, Linas Baltrunas, and Domonkos Tikk.
\newblock Session-based recommendations with recurrent neural networks.
\newblock {\em arXiv preprint arXiv:1511.06939}, 2015.

\bibitem[\protect\citeauthoryear{Hu \bgroup \em et al.\egroup }{2022}]{hu2022membership}
Hongsheng Hu, Zoran Salcic, Lichao Sun, Gillian Dobbie, Philip~S Yu, and Xuyun Zhang.
\newblock Membership inference attacks on machine learning: A survey.
\newblock {\em ACM Computing Surveys (CSUR)}, 54(11s):1--37, 2022.

\bibitem[\protect\citeauthoryear{Koren \bgroup \em et al.\egroup }{2009}]{koren2009matrix}
Yehuda Koren, Robert Bell, and Chris Volinsky.
\newblock Matrix factorization techniques for recommender systems.
\newblock {\em Computer}, 42(8):30--37, 2009.

\bibitem[\protect\citeauthoryear{Liu \bgroup \em et al.\egroup }{2018}]{liu2018stamp}
Qiao Liu, Yifu Zeng, Refuoe Mokhosi, and Haibin Zhang.
\newblock Stamp: short-term attention/memory priority model for session-based recommendation.
\newblock In {\em KDD}, pages 1831--1839, 2018.

\bibitem[\protect\citeauthoryear{McAuley \bgroup \em et al.\egroup }{2015}]{mcauley2015image}
Julian McAuley, Christopher Targett, Qinfeng Shi, and Anton Van Den~Hengel.
\newblock Image-based recommendations on styles and substitutes.
\newblock In {\em SIGIR}, pages 43--52, 2015.

\bibitem[\protect\citeauthoryear{Narducci \bgroup \em et al.\egroup }{2015}]{narducci2015recommender}
Fedelucio Narducci, Cataldo Musto, Marco Polignano, Marco de~Gemmis, Pasquale Lops, and Giovanni Semeraro.
\newblock A recommender system for connecting patients to the right doctors in the healthnet social network.
\newblock In {\em WWW}, pages 81--82, 2015.

\bibitem[\protect\citeauthoryear{Pardau}{2018}]{pardau2018california}
Stuart~L Pardau.
\newblock The california consumer privacy act: Towards a european-style privacy regime in the united states.
\newblock {\em J. Tech. L. \& Pol'y}, 23:68, 2018.

\bibitem[\protect\citeauthoryear{Ribeiro \bgroup \em et al.\egroup }{2015}]{ribeiro2015mlaas}
Mauro Ribeiro, Katarina Grolinger, and Miriam~AM Capretz.
\newblock Mlaas: Machine learning as a service.
\newblock In {\em ICMLA}, pages 896--902, 2015.

\bibitem[\protect\citeauthoryear{Rosen}{2011}]{rosen2011right}
Jeffrey Rosen.
\newblock The right to be forgotten.
\newblock {\em Stanford Law Review}, 64:88, 2011.

\bibitem[\protect\citeauthoryear{Salem \bgroup \em et al.\egroup }{2019}]{salem2019ml}
Ahmed Salem, Yang Zhang, Mathias Humbert, Mario Fritz, and Michael Backes.
\newblock Ml-leaks: Model and data independent membership inference attacks and defenses on machine learning models.
\newblock In {\em NDSS Symposium}. Internet Society, 2019.

\bibitem[\protect\citeauthoryear{Sarwar \bgroup \em et al.\egroup }{2001}]{sarwar2001item}
Badrul Sarwar, George Karypis, Joseph Konstan, and John Riedl.
\newblock Item-based collaborative filtering recommendation algorithms.
\newblock In {\em WWW}, pages 285--295, 2001.

\bibitem[\protect\citeauthoryear{Sedhain \bgroup \em et al.\egroup }{2014}]{sedhain2014social}
Suvash Sedhain, Scott Sanner, Darius Braziunas, Lexing Xie, and Jordan Christensen.
\newblock Social collaborative filtering for cold-start recommendations.
\newblock In {\em RecSys}, pages 345--348, 2014.

\bibitem[\protect\citeauthoryear{Sedhain \bgroup \em et al.\egroup }{2015}]{sedhain2015autorec}
Suvash Sedhain, Aditya~Krishna Menon, Scott Sanner, and Lexing Xie.
\newblock Autorec: Autoencoders meet collaborative filtering.
\newblock In {\em WWW}, pages 111--112, 2015.

\bibitem[\protect\citeauthoryear{Sedhain \bgroup \em et al.\egroup }{2017}]{sedhain2017low}
Suvash Sedhain, Aditya Menon, Scott Sanner, Lexing Xie, and Darius Braziunas.
\newblock Low-rank linear cold-start recommendation from social data.
\newblock In {\em AAAI}, volume~31, 2017.

\bibitem[\protect\citeauthoryear{Shokri \bgroup \em et al.\egroup }{2017}]{shokri2017membership}
Reza Shokri, Marco Stronati, Congzheng Song, and Vitaly Shmatikov.
\newblock Membership inference attacks against machine learning models.
\newblock In {\em S\&P}, pages 3--18. IEEE, 2017.

\bibitem[\protect\citeauthoryear{Song and Mittal}{2021}]{song2021systematic}
Liwei Song and Prateek Mittal.
\newblock Systematic evaluation of privacy risks of machine learning models.
\newblock In {\em USENIX Security}, pages 2615--2632, 2021.

\bibitem[\protect\citeauthoryear{Song and Shmatikov}{2019}]{song2019auditing}
Congzheng Song and Vitaly Shmatikov.
\newblock Auditing data provenance in text-generation models.
\newblock In {\em KDD}, pages 196--206, 2019.

\bibitem[\protect\citeauthoryear{Sun \bgroup \em et al.\egroup }{2019}]{sun2019bert4rec}
Fei Sun, Jun Liu, Jian Wu, Changhua Pei, Xiao Lin, Wenwu Ou, and Peng Jiang.
\newblock Bert4rec: Sequential recommendation with bidirectional encoder representations from transformer.
\newblock In {\em CIKM}, pages 1441--1450, 2019.

\bibitem[\protect\citeauthoryear{Tang and Wang}{2018}]{tang2018personalized}
Jiaxi Tang and Ke~Wang.
\newblock Personalized top-n sequential recommendation via convolutional sequence embedding.
\newblock In {\em WSDM}, pages 565--573, 2018.

\bibitem[\protect\citeauthoryear{Tang \bgroup \em et al.\egroup }{2016}]{tang2016recommendations}
Jiliang Tang, Charu Aggarwal, and Huan Liu.
\newblock Recommendations in signed social networks.
\newblock In {\em WWW}, pages 31--40, 2016.

\bibitem[\protect\citeauthoryear{Wang \bgroup \em et al.\egroup }{2022}]{wang2022debiasing}
Zihan Wang, Na~Huang, Fei Sun, Pengjie Ren, Zhumin Chen, Hengliang Luo, Maarten de~Rijke, and Zhaochun Ren.
\newblock Debiasing learning for membership inference attacks against recommender systems.
\newblock In {\em KDD}, pages 1959--1968, 2022.

\bibitem[\protect\citeauthoryear{Wu \bgroup \em et al.\egroup }{2019}]{wu2019neural}
Le~Wu, Peijie Sun, Yanjie Fu, Richang Hong, Xiting Wang, and Meng Wang.
\newblock A neural influence diffusion model for social recommendation.
\newblock In {\em SIGIR}, pages 235--244, 2019.

\bibitem[\protect\citeauthoryear{Ye \bgroup \em et al.\egroup }{2022}]{ye2022enhanced}
Jiayuan Ye, Aadyaa Maddi, Sasi~Kumar Murakonda, Vincent Bindschaedler, and Reza Shokri.
\newblock Enhanced membership inference attacks against machine learning models.
\newblock In {\em CCS}, pages 3093--3106, 2022.

\bibitem[\protect\citeauthoryear{Yeom \bgroup \em et al.\egroup }{2018}]{yeom2018privacy}
Samuel Yeom, Irene Giacomelli, Matt Fredrikson, and Somesh Jha.
\newblock Privacy risk in machine learning: Analyzing the connection to overfitting.
\newblock In {\em CSF}, pages 268--282. IEEE, 2018.

\bibitem[\protect\citeauthoryear{Zhang \bgroup \em et al.\egroup }{2019}]{zhang2019deep}
Shuai Zhang, Lina Yao, Aixin Sun, and Yi~Tay.
\newblock Deep learning based recommender system: A survey and new perspectives.
\newblock {\em ACM Computing Surveys (CSUR)}, 52(1):1--38, 2019.

\bibitem[\protect\citeauthoryear{Zhang \bgroup \em et al.\egroup }{2021}]{zhang2021membership}
Minxing Zhang, Zhaochun Ren, Zihan Wang, Pengjie Ren, Zhunmin Chen, Pengfei Hu, and Yang Zhang.
\newblock Membership inference attacks against recommender systems.
\newblock In {\em CCS}, pages 864--879, 2021.

\bibitem[\protect\citeauthoryear{Zhou \bgroup \em et al.\egroup }{2018}]{zhou2018micro}
Meizi Zhou, Zhuoye Ding, Jiliang Tang, and Dawei Yin.
\newblock Micro behaviors: A new perspective in e-commerce recommender systems.
\newblock In {\em WSDM}, pages 727--735, 2018.

\bibitem[\protect\citeauthoryear{Zhou \bgroup \em et al.\egroup }{2019}]{zhou2019deep}
Guorui Zhou, Na~Mou, Ying Fan, Qi~Pi, Weijie Bian, Chang Zhou, Xiaoqiang Zhu, and Kun Gai.
\newblock Deep interest evolution network for click-through rate prediction.
\newblock In {\em AAAI}, volume~33, pages 5941--5948, 2019.

\bibitem[\protect\citeauthoryear{Zhu \bgroup \em et al.\egroup }{2023}]{zhu2023membership}
Zhihao Zhu, Chenwang Wu, Rui Fan, Defu Lian, and Enhong Chen.
\newblock Membership inference attacks against sequential recommender systems.
\newblock In {\em WWW}, pages 1208--1219, 2023.

\end{thebibliography}

\clearpage

\section*{Appendix}
\renewcommand\thesubsection{A. 1}
\subsection{Details of Datasets and Baseline Settings.}\label{appendix::dataset}

\paragraph{Dataset Description.} To reduce the adverse effect of data sparsity on the performance of recommender systems~\cite{he2017neural}, following the previous work~\cite{wang2022debiasing}, we remove users and items with less than 5 interaction records for all three datasets. While the user-item interactions from three datasets are explicit feedback data, we focus on implicit task of recommender systems and assign value of 1 to user-item pairs if there are interactions, otherwise 0~\cite{ding2018improving,he2016fast}.

\paragraph{Dataset Partition.} To ensure a fair comparison with baseline attack methods~\cite{zhang2021membership,wang2022debiasing}, we follow Zhang et al.~\cite{zhang2021membership} to split each dataset into three disjoint subsets, i.e., a shadow dataset, a target dataset, and a dataset for extracting item features. For baselines, we follow their workflow to conduct experiments to reproduce results. For our proposed method, only target dataset and dataset for extracting item features are utilized in our experiments. In Section~\ref{sec:effectiveness}, we show that with less information in hand, the attacker can achieve far better attack performance than the baselines using our proposed SF-MIAs.

\paragraph{Parameter Settings for Different Recommender Systems.} In our experiments, we follow existing works~\cite{zhang2021membership,wang2022debiasing} to select recommender systems. We utilize various recommender systems including ICF, NCF, BERT4Rec, Caser, and GRU4Rec to evaluate attack performance of proposed shadow-free MIAs. We list the settings of main parameters existing in these recommender systems in Table~\ref{tab:recsetting}. We follow the standard training recipes to train the recommenders (source code link provided in~\ref{appendix::repro}), exemplified by the BERT4Rec model. As shown in Table 5 in Appendix, we set the batch size as 128 and the learning rate as 0.001, and use Adam as the optimizer. For example, model trained on MovieLens-1M has a training and testing NDCG@10 0.983 and 0.981, demonstrating high generalizability achieved in the model.

\paragraph{Baseline Details.} ST-MIA~\cite{zhang2021membership} is the first algorithm proposed to implement MIA against recommender systems, following the pipeline of shadow training described in Section~\ref{sec:shadow-training}. In this method, attacker can utilize shadow training technique to build a shadow recommender system, in order to mimic the behavior of target recommender system. Attacker utilizes the difference between users' recommendations and their historical interactions to determine whether or not a user is member. Lower difference indicates that a user's data is more likely to be used to train target recommender system. Specifically, attacker designs an attack model which is a binary classifier and utilizes shadow recommender to build labelled training dataset for the attack model, since the ground truth membership status of users in shadow recommender is already known by the attaker. DL-MIA~\cite{wang2022debiasing} also follows shadow training pipeline and further improves ST-MIA through debias learning. In this work, authors claim that MIA against recommender systems is biased when attacker has no knowledge about data distribution of target dataset and model architecture of target recommender system. As a result, they propose a variational auto-encoder based disentangled encoder to reduce the gap between shadow recommender system and target recommender system. In addition, items' latent feature vectors are not observable for attacker when training target recommender systems, which would make users' feature vectors inaccurate. Thus, they propose to utilize a weight estimator to reduce the estimation bias.   

\paragraph{Attack Settings.} Table~\ref{tab:notationBaselines} gives a list of illustrations of all 2-letter combinations for two baselines in our experiments. We use these combinations to represent different configurations for simplicity purposes. For example, ``BT'' represents that the attacker in baseline methods uses BERT4Rec as a shadow recommender and Ta-feng as a shadow dataset. ``IM'' represents that the attacker in baseline methods uses ICF as a shadow recommender and MovieLens-1M as a shadow dataset.

\begin{table}

\centering
\resizebox{\linewidth}{!}{\begin{tabular}{cl}
\toprule 
\textbf{Recommender system} & \textbf{Settings} \\

 \midrule
\multirow{1}{*}{ICF} & Cosine similarity metric\\

 \midrule
\multirow{3}{*}{NCF} &  Embedding size= 16, batch size= 256, Adam optimizer,  \\ 
& hidden unit= (128, 64, 32), learning rate= 0.001 \\
\midrule
\multirow{3}{*}{BERT4Rec} & Batch size= 128, Adam optimizer, dropout= 0.1, \\ 
& hidden size = 256, learning rate= 0.001\\
\midrule
\multirow{3}{*}{Caser} & Embedding size= 50, batch size= 512, Adam optimizer,  \\
& dropout= 0.5, learning rate= 0.001\\
\midrule
\multirow{3}{*}{GRU4Rec} &   Batch size= 50, Adagrad optimizer, dropout= 0.5,  \\ 
& hidden size= 100, learning rate= 0.01, momentum= 0\\

\bottomrule
\end{tabular}}
\caption{Parameter settings for different recommender systems. }
\label{tab:recsetting}
\end{table}

\begin{table}

\centering
\resizebox{\linewidth}{!}{\begin{tabular}{cl}
\toprule 
\textbf{2-letter Combinations} & \textbf{Illustrations} \\

 \midrule
\multirow{3}{*}{BT} & Shadow recommender is implemented by \\
 & BERT4Rec on Ta-feng dataset.\\
 \midrule
\multirow{3}{*}{NT} & Shadow recommender is implemented by \\ 
& NCF on Ta-feng dataset.\\
\midrule
\multirow{3}{*}{IT} & Shadow recommender is implemented by \\ 
& ICF on Ta-feng dataset.\\
\midrule
\multirow{3}{*}{GA} & Shadow recommender is implemented by \\ 
& GRU4Rec on Amazon Beauty dataset.\\
\midrule
\multirow{3}{*}{CA} & Shadow recommender is implemented by \\ 
& Caser on Amazon Beauty dataset.\\
\midrule
\multirow{3}{*}{BA} & Shadow recommender is implemented by \\ 
& BERT4Rec on Amazon Beauty dataset.\\
\midrule
\multirow{3}{*}{BM} & Shadow recommender is implemented by \\ 
& BERT4Rec on MovieLens-1M dataset.\\
\midrule
\multirow{3}{*}{IM} & Shadow recommender is implemented by \\ 
& ICF on MovieLens-1M dataset.\\
\midrule
\multirow{3}{*}{CM} & Shadow recommender is implemented by \\ 
& Caser on MovieLens-1M dataset.\\
\midrule
\multirow{3}{*}{GM} & Shadow recommender is implemented by \\ 
& GRU4Rec on MovieLens-1M dataset.\\
\midrule
\multirow{3}{*}{NM} & Shadow recommender is implemented by \\ 
& NCF on MovieLens-1M dataset.\\

\bottomrule
\end{tabular}}
\caption{Notations of attack settings for two baselines.}
\label{tab:notationBaselines}
\end{table}

\paragraph{Experiments Environment.} We implement the experiments using Python 3.9 with NVIDIA V100 GPUs on NCI high-performance computing GADI.

\renewcommand\thesubsection{A. 2}
\subsection{Reproducibility.}\label{appendix::repro}
To facilitate the reproducibility of the results reported in this paper, code and data used in this work is provided at \url{https://github.com/XiaoxiaoChi-code/shadow-free-MIAs.git}. For the implementation of baselines, we use the code they provided in their papers~\cite{zhang2021membership,wang2022debiasing}.

\renewcommand\thesubsection{A. 3}
\subsection{Additional Experimental Results.}\label{appendix::results}

\begin{figure}
    \centering
    \includegraphics[width=0.8\linewidth]{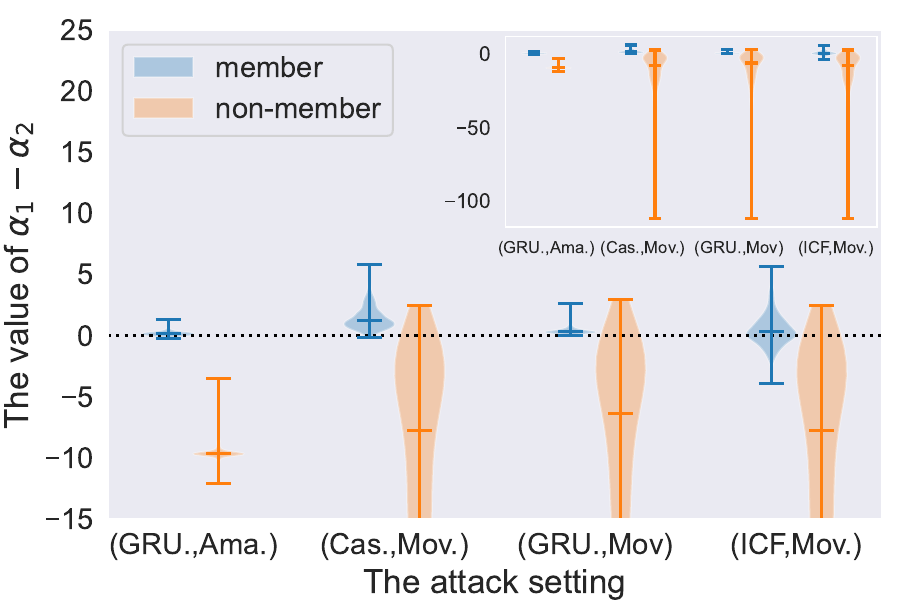}
   \caption{Visualization of $\alpha_1-\alpha_2$ data distribution. As we can see, the distributions of member and non-member users are very different.}
   \label{fig:violin}
\end{figure}

\paragraph{$\alpha_1 - \alpha_2$ Distribution.}
The experimental quantity results on understanding why the proposed attack work are provided in Figure~\ref{fig:violin}. It shows a visualization of $\alpha_1 - \alpha_2$ for member users and non-member users. From this visualization, we can see that the distributions of member and non-member users are very different, explaining why our proposed attacks are highly effective.

\begin{figure}[t!]
    \centering
  \subfloat[Understanding impact of $n$.]{%
       \includegraphics[width=0.23\textwidth]{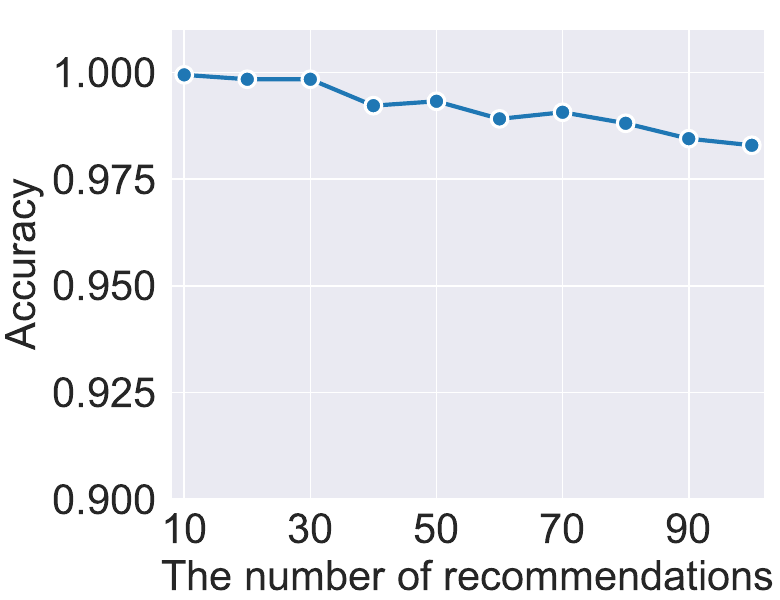}}
       \vspace{1.5pt}
  \subfloat[Understanding impact of $l$.]{%
        \includegraphics[width=0.23\textwidth]{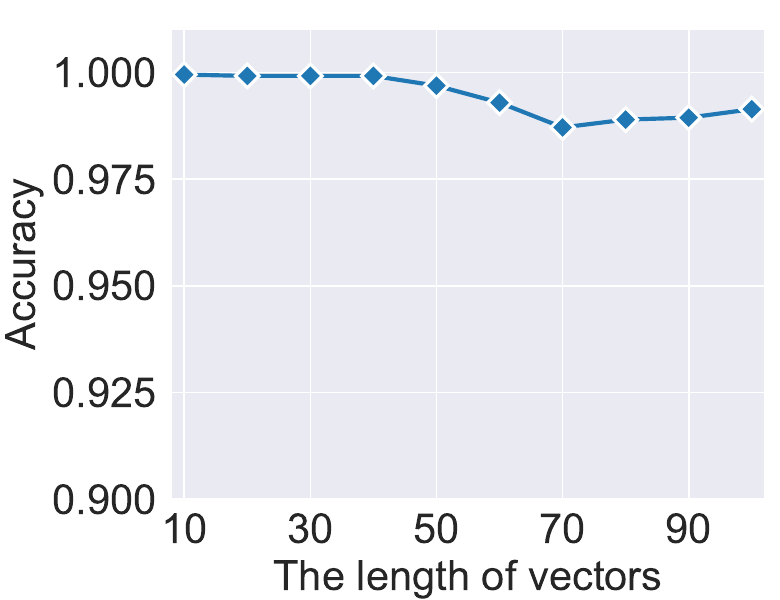}}
  \caption{Understanding the impact of the number of recommendations and the length of the feature vector in shadow-free MIAs. As we can see, shadow-free MIAs are stable when varying the two factors.}
  \label{fig:overun}
\end{figure}

\paragraph{Impact of Parameter $n,l$ on the Attack Performance.} We study how the number of recommendations $n$ and the length of vectors $l$ can influence the attack performance of the proposed attack. Figure~\ref{fig:overun} shows the shadow-free MIAs are stable and effective under different attack settings. The number of recommendations and the length of item feature vectors only slightly influence the attack performance.

\end{document}